\documentclass{article}
\usepackage[totalwidth=13.0cm,totalheight=20.0cm]{geometry}

\usepackage{latexsym,amsthm,amsmath,amssymb,url}

\newcommand{\HS}{\textsc{HitSet}}

\newcommand{\HSnk}{\textsc{HitSet}$(n-k,k)$}
\newcommand{\HSnkp}{\textsc{HitSet}$(n-k,k+d)$}
\newcommand{\HSmk}{\textsc{HitSet}$(m-k,k)$}

\newcommand{\HSmkk}{\textsc{HitSet}$(p,m+p)$}

\newtheorem{lemma}{Lemma}
\newtheorem{corollary}{Corollary}
\newtheorem{definition}{Definition}
\newtheorem{theorem}{Theorem}
\newtheorem{krule}{Reduction Rule}

\begin{document}

\title{Kernels for Below-Upper-Bound Parameterizations of the Hitting Set and Directed Dominating Set Problems}

\author{
 G. Gutin, M. Jones and A. Yeo\\
{\small Royal Holloway, University of London}\\[-3pt]
{\small Egham, Surrey, TW20 0EX, UK}\\[-3pt]
{\small \url{{gutin|markj|anders}@cs.rhul.ac.uk}}\\
}
\date{}
\maketitle

\begin{abstract}
In the {\sc Hitting Set} problem, we are given a collection $\cal F$ of subsets of a ground set $V$ and an integer $p$,
and asked whether $V$ has a $p$-element subset that intersects each set in $\cal F$.
We consider two parameterizations of {\sc Hitting Set} below tight upper bounds: $p=m-k$ and $p=n-k$.
In both cases $k$ is the parameter. We prove that the first parameterization is fixed-parameter tractable,
but has no polynomial kernel unless coNP$\subseteq$NP/poly.
The second parameterization is W[1]-complete, but the introduction of an additional parameter, the degeneracy of
the hypergraph $H=(V,{\cal F})$, makes the problem not only fixed-parameter tractable,
but  also one with a linear kernel. Here the degeneracy of $H=(V,{\cal F})$ is the minimum
integer $d$ such that for each $X\subset V$ the hypergraph with vertex set $V\setminus X$ and edge set containing all edges
of $\cal F$ without vertices in $X$, has a vertex of degree at most $d.$

In {\sc Nonblocker} ({\sc Directed Nonblocker}), we are given an undirected graph (a directed graph) $G$ on $n$ vertices and an integer $k$,
and asked whether $G$ has a set $X$ of $n-k$ vertices such that for each vertex $y\not\in X$ there is an edge (arc) from
a vertex in $X$ to $y$.
{\sc Nonblocker} can be viewed as a special case of {\sc Directed Nonblocker} (replace an undirected graph by a symmetric digraph).
Dehne et al. (Proc. SOFSEM 2006) proved that {\sc Nonblocker} has a linear-order kernel. We obtain a linear-order kernel for
{\sc Directed Nonblocker}.
\end{abstract}

\section{Introduction, Terminology and Notation}\label{sec:i}

In the {\sc Hitting Set} problem, we are given a collection $\cal F$ of subsets of a ground set $V$ and an integer $p$,
and asked whether $V$ has a $p$-element subset that intersects each set in $\cal F$.
It is a well-known problem with various applications, e.g., in software testing \cite{JH2003}, in computer networks
\cite{KRWWZ} and in bioinformatics \cite{RS2002}.  {\sc Hitting Set} is equivalent to the Set Cover
problem and several of its special cases are of importance (e.g., the Vertex Cover and Dominating Set problems).
{\sc Hitting Set} is NP-complete and its standard parameterization (when $p$ is the parameter) is W[2]-complete.
(We provide basic parameterized complexity terminology and notation in Subsection \ref{sec:pc}.)
A few alternative parameterizations of {\sc Hitting Set} have also been studied and we briefly overview them below.
To facilitate our discussion of various parameterizations of {\sc Hitting Set}, consider the
following generic parameterization:

 \begin{quote}
{\HS}($p$,$\kappa$)\\ \nopagebreak
  \emph{Instance:} A set $V$, a collection $\cal F$ of subsets of $V$.\\
    \nopagebreak
  \emph{Parameter:} $\kappa$.\\ \nopagebreak
  \emph{Question:} Does $(V,{\cal F})$ have a hitting set $S$ of size at most $p$? (A subset $S$ of $V$ is called a {\em hitting set} if $S\cap F\neq \emptyset$
  for each $F\in {\cal F}$.)
 \end{quote}
In what follows, $n$ stands for the size of $V$ and $m$ for the size of $\cal F$.\\

Aside
from {\HS}($p$,$p$), the standard parameterization  of  {\sc Hitting Set}, the most well-known parameterization is
{\HS}($p$,$p+s$), where $s$
is the maximum size of a set in $\cal F$.
This parameterization is fixed-parameter tractable and has a kernel of
size at most $s^p$
(see Downey and Fellows \cite{DowneyFellows99}).
Using the Sunflower Lemma, Flum and Grohe \cite{FlumGrohe06} obtained a
kernel of size $O(sp^ss!)$.
Abu-Khzam \cite{Abu-Khzam10} recently proved that {\HS}($p$,$p+s$) has a
kernel in which the number of
elements in the ground set $V$ is at most $(2s-1)p^{s-1}+p.$ Dom et al.
\cite{DLS09} proved that {\HS}($p$,$p+s$)
does not admit a polynomial-size kernel unless coNP$\subseteq$NP/poly. Dom
et al. \cite{DLS09} also proved that
{\HS}($p$,$p+m$) and {\HS}($p$,$p+n$)
have exponential-size kernels but no polynomial-size kernels
unless coNP$\subseteq$NP/poly.

In this paper, we study two parameterizations: \HSmk{} and \HSnk, as well as \HSnk{} with an additional parameter.
Both parameterizations are of the type "below a tight upper bound." Indeed, both $m$ and $n$ are
tight upper bounds as it is easy to see that there is always a hitting set of size at most $m$ ($n$,
respectively) and to
construct an infinite family of instances of {\sc Hitting Set} in which no hitting set is of size less
than $m$ ($n$, respectively). A brief overview is given in Subsection \ref{sec:bb} of some well-known results
on problems parameterized below tight upper bounds.  Subsection \ref{sec:t} is devoted to hypergraph terminology and notation;
note that some terminology and notation that we use is new or nonstandard. A very brief introduction to fixed-parameter algorithmics
is given in Subsection \ref{sec:pc}.

In Section \ref{sec:m}, we prove that \HSmk{} has a kernel with at most $k4^k$  elements of $V$ and at most $k4^k$  sets. In our proof,
we use a technique called greedy localization, introduced by Chen et al. \cite{CFJK2004}.
This technique is often compared to the well-known iterative compression, see, e.g., Dehne et al. \cite{DFRS2004}.
We also prove that \HSmk{} has no kernel of polynomial size unless coNP$\subseteq$NP/poly.
In our proof, we use the result of Dom et al. \cite{DLS09} on
{\HS}($p$,$p+m$) mentioned above.

In the problem {\sc Nonblocker},
we are given a graph $G=(V,E)$
and an integer $k$, and asked whether there is a set $X\subseteq V$ of size at most $|V|-k$ such that each
vertex $v\in V\setminus X$ is adjacent to a vertex in $X.$ Here $k$ is the parameter. Note that {\sc Nonblocker} is
a below-tight-upper-bound parameterization of the {\sc Dominating Set} problem.
It is well-known that {\sc Nonblocker} can be reduced
to {\HS}($n-k$,$k$) (see, e.g., \cite{FlumGrohe06}, p. 18) and, thus,
our no-polynomial-kernel result is in a
sharp contrast to a linear-order-kernel result of Dehne et al. \cite{DFFPR2006} for
{\sc Nonblocker}.

In Section \ref{sec:n}, we show that \HSnk{} is W[1]-complete, but the introduction of the second parameter,
the corresponding hypergraph degeneracy (defined in Subsection \ref{sec:t}), makes the problem not only
fixed-parameter tractable, but also  one with a linear kernel. Each hypergraph of maximum degree
$d$ is $d$-degenerate, but the family of $d$-degenerate hypergraphs has its maximum degree unbounded by any function of $d$.
Thus, our result is an extension of the corresponding result when the maximum degree is the additional parameter.
The {\sc Directed Nonblocker} problem is an extension of {\sc Nonblocker} to directed graphs: we are given a
directed graph $G=(V,A)$ and an integer $k$, and asked whether there is a set $X\subseteq V$ of size at most
$|V|-k$ such that for each vertex $v\in V\setminus X$ there is an arc from a vertex of $X$ to $v.$
Using our polynomial-size kernel result for \HSnk{} with the additional
parameter, we show that {\sc Directed Nonblocker} has a kernel with at most $k^2+k-1$ vertices.

In Section \ref{sec:nb}, we improve the last result by showing that {\sc Directed Nonblocker} has a kernel with at most $3k-1$ vertices.
To prove this result we use an inequality for the domination number of a digraph with at most one vertex of in-degree zero and no isolated vertices.
Further research is discussed in Section \ref{sec:fr}.

\subsection{Problems Parameterized below Tight Upper Bounds}\label{sec:bb}

Mahajan and Raman \cite{Mahajan99} were the first to recognize both practical and theoretical importance of parameterizing problems
above tight lower bounds or below tight upper bounds. Further arguments for the importance of parameterizing problems
above and below tight bounds were given in \cite{MahajanRamanSikdar09,Niedermeier06}.
One example of a problem parameterized below an upper bound is {\sc Maximum Clique} parameterized below $n$,
the number of vertices in the input graph. Unlike the standard parameterization of {\sc Maximum Clique} which is W[1]-complete,
the parameterization below $n$ is fixed-parameter tractable (and has a linear-order kernel) simply because it is equivalent to the
standard parameterization of {\sc Vertex Cover}. The parameterization below $n$ is important in bioinformatics applications,
where the maximum order of a clique is close to $n$ \cite{ACFLSS2004}.

However, in many cases establishing parameterized complexity of a problem parameterized below a tight upper bound is less straightforward
and has been stated as an open question. One such well-known problem is {\sc Directed Feedback Vertex Set}: given a digraph $D$ and an integer $k$, decide whether
$D$ has an acyclic induced subgraph on at least $n-k$ vertices. The parameterized complexity of {\sc Directed Feedback Vertex Set}
was a long standing open question solved by Chen et al. \cite{CLLOR2008} who established its fixed-parameter tractability. Other well-known examples are
{\sc Bipartization} (decide whether a graph has a bipartite induced subgraph on $n-k$ vertices) which was proved to be fixed-parameter tractable by Reed et al. \cite{RSV2004} and {\sc Almost 2-Sat} (decide whether there is a truth assignment that satisfies at least $m-k$ clauses in a 2-CNF formula with $m$ clauses) which was proved to be fixed-parameter tractable by Razgon and O'Sullivan \cite{RO2009}. Interestingly, no polynomial-size kernel is known for any of the three problems and it is still unknown whether such a kernel exists.

Certainly, not every natural problem parameterized below a tight upper bound is fixed-parameter tractable.
A trivial example of such a problem is {\sc Almost 3-Sat} (decide whether there is a truth assignment that satisfies at
least $m-k$ clauses in a 3-CNF formula with $m$ clauses). A less trivial example is the following problem:
given a graph $G$, its maximal matching $M$ and an integer $k$, decide whether $G$ has a vertex cover with at most
$2|M|-k$ vertices. This problem was proved to be W[1]-hard by Gutin et al. \cite{GKLM2010}.
(Here we assume that W[1]$\neq$FPT as widely believed.)


\subsection{Hypergraphs}\label{sec:t}

While studying {\sc Hitting Set}, it will be more convenient for us to use hypergraph terminology and notation, we
introduce the relevant terminology and notation in this subsection.

A {\em hypergraph} $H = (V, {\cal F})$ consists of a nonempty set $V$ of {\em vertices} and a family $\cal F$ of nonempty subsets of
$V$ called {\em edges} of $H.$ Note that $\cal F$ may have {\em parallel} edges, i.e., copies of the same subset
of $V.$ For any vertex $v \in V$, and any ${\cal E} \subseteq {\cal F}$, ${\cal E}[v]$ is the set of edges in
${\cal E}$ containing $v$, $N[v]$ is the set of all vertices contained in edges of ${\cal F}[v]$, and
the {\em degree} of $v$ is $d(v) = |{\cal F}[v]|$.  For a subset $T$ of vertices, ${\cal F}[T]=\bigcup_{v\in T}{\cal F}[v].$

 {\em Deleting} an edge $e$ from a {\em hypergraph}
$H = (V, {\cal F})$ results in a new hypergraph $H-e$ with vertex set $V$ and edge set ${\cal F}\setminus \{e\}$.
{\em Deleting} a vertex $v$ from a {\em hypergraph}
$H = (V, {\cal F})$ results in a new hypergraph $H-v$ with vertex set $V\setminus \{v\}$ and edge set
$\{e\setminus  \{v\}:\ e\in {\cal F}\}.$

A set $T$ of vertices {\em hits} all edges in ${\cal F}[T]$ and an edge $e$ is {\em hit} by any vertex belonging to it.
A set $S \subseteq V$ is called  a \emph{hitting set} of a hypergraph $H=(V, {\cal F})$ if it hits ${\cal F}$.
{\sc Hitting Set} can be formulated as a problem in which we are given a hypergraph $H$ and an integer $p$ and asked
whether $H$ contains a hitting set of size at most $p.$

For a hypergraph $H = (V, {\cal F})$ and a set $X\subset V$, the subhypergraph $H\circleddash X$ is obtained from $H$ by deleting the set $\cal E$ of all edges hit by $X$ and all vertices contained only in $\cal E$.
A hypergraph $H = (V, {\cal F})$ is \emph{$d$-degenerate} if, for all $X \subset V$, the subhypergraph $H\circleddash X$ contains a vertex of degree at most $d$. The \emph{degeneracy} ${\rm deg}(H)$ of a hypergraph $H$ is the smallest $d$ for which $H$ is $d$-degenerate.

The degeneracy of a hypergraph can be calculated in linear time using the following algorithm.
Pick a vertex $v_1$ in $H$ of minimum degree $d_1$, and  set $H:=H\circleddash {\{v_1\}}$.
Pick a vertex $v_2$ of minimum degree $d_2$  and set $H:=H\circleddash {\{v_2\}}$, and so on. Then $d = \max \{ d_i: i \in [n] \}$
is the degeneracy of $H$. (It is clear that the degeneracy of $H$ must be at least $d$;
the equality follows by observing that for any $X \subset V$ the smallest numbered vertex $u_i\in V\setminus X$
has degree at most $d_i$ in $H\circleddash X$.)

\subsection{Fixed-Parameter Tractability and Kernels}\label{sec:pc}

A \emph{parameterized problem} is a subset $L\subseteq \Sigma^* \times
\mathbb{N}$ over a finite alphabet $\Sigma$. $L$ is
\emph{fixed-parameter tractable} if the membership of an instance
$(I,k)$ in $\Sigma^* \times \mathbb{N}$ can be decided in time
$f(k)|I|^{O(1)}$ where $f$ is a computable function of the
{\em parameter} $k$ only~\cite{DowneyFellows99,FlumGrohe06,Niedermeier06}.
Given a parameterized problem $L$,
a \emph{kernelization of $L$} is a polynomial-time
algorithm that maps an instance $(x,k)$ to an instance $(x',k')$ (the
\emph{kernel}) such that (i)~$(x,k)\in L$ if and only if
$(x',k')\in L$, (ii)~ $k'\leq h(k)$, and (iii)~$|x'|\leq g(k)$ for some
functions $h$ and $g$. The function $g(k)$ is called the {\em size} of the kernel.
It is well-known \cite{DowneyFellows99,FlumGrohe06,Niedermeier06} that a decidable parameterized problem $L$ is fixed-parameter
tractable if and only if it has a kernel. Polynomial-size kernels are of
main interest, due to applications \cite{DowneyFellows99,FlumGrohe06,Niedermeier06}, but unfortunately not all fixed-parameter problems
have such kernels unless  coNP$\subseteq$NP/poly, see, e.g., \cite{BDFH09,BTY09,DLS09}.

\section{Hitting Set parameterized below $m$}\label{sec:m}

In this section we consider \HSmk.  Let $H = (V, {\cal F})$ be a hypergraph.

We begin with some reduction rules. The first two reduction rules have been used by previous \textsc{Hitting Set} algorithms \cite{Abu-Khzam10}, \cite{Weihe98}. The third is a trivial rule included to simplify later proofs.

\begin{krule}\label{subset2}
 If there exist distinct $e,e' \in {\cal F}$ such that $e \subseteq e'$,  set $H:=H-e$ and $k:=k-1$.
\end{krule}

\begin{krule}\label{subelement2}
  If there exist $u,v \in V$ such that $u \neq v$ and ${\cal F}[u] \subseteq {\cal F}[v]$,   set $H:=H-u$.
\end{krule}

\begin{krule}\label{trivial2}
If there exist $v \in V$, $e \in {\cal F}$ such that ${\cal F}[v] = \{e\}$ and $e = \{ v \}$, then delete $v$ and $e$.
\end{krule}

\begin{lemma}\label{mreduction}
 Let $(H, k)$ and $(H', k')$ be instances of \HSmk{} such that $(H', k')$ is derived from $(H, k)$ by repeated applications of Rules \ref{subset2}, \ref{subelement2} and \ref{trivial2}. Then $(H', k')$ is a {\sc Yes}-instance if and only if $(H, k)$ is a {\sc Yes}-instance.
\end{lemma}

\begin{proof}
Rule \ref{subset2}: Any vertex in $V$ which hits $e$ will also hit $e'$. Therefore a set $S \subseteq V$ is a hitting set for $H$ if and only if it is also a hitting set for $H - e$. In $H - e$ the difference between $m$ and the size of the desired hitting set is one less, so we reduce $k$ by $1$.

Rule \ref{subelement2}: Any edge which is hit by $u$ is also hit by $v$. Therefore, if $S$ is a hitting set containing $u$, we can get another hitting set of equal size or smaller by removing $u$ and adding $v$. Therefore we may assume $u$ is not in the hitting set
 and  delete $u$ from $H$.

Rule \ref{trivial2}: The proof is trivial.
\end{proof}

\begin{lemma}\label{noIslands2}
 Let $(H = (V, {\cal F}), k)$ be an instance of \HSmk{} which is reduced by Rules \ref{subset2}, \ref{subelement2} and \ref{trivial2} and ${\cal F}\neq \emptyset$. Then for all $v \in V$, $d(v) \ge 2$, and for all $e \in {\cal F}$, $|e| \ge 2$.
\end{lemma}
\begin{proof}
Consider $v \in V$. Suppose $d(v) = 0$. Then trivially, ${\cal F}[v] \subseteq {\cal F}[u]$ for any $u \in V$, and so Rule \ref{subelement2} applies, a contradiction. Suppose $d(v) = 1$. Then let $e$ be the single edge containing $v$. Either $e$ contains another vertex $u$, in which case ${\cal F}[v] \subseteq {\cal F}[u]$ and Rule \ref{subelement2} applies,
or $e = \{ v \}$, in which case Rule \ref{trivial2} applies, a contradiction. Thus, $d(v) \ge 2$.
A similar argument, using Rule \ref{subset2} instead of Rule \ref{subelement2}, can be used to show that $|e| \ge 2$ for all $e \in {\cal F}$.
\end{proof}

We now introduce the concept of a \emph{mini-hitting set}. Lemma \ref{miniDominatingSet} shows that the problem of finding a hitting set of size $m-k$ is equivalent to the problem of finding a mini-hitting set.

\begin{definition}
 A \emph{mini-hitting set} is a set $S_{\textsc{mini}} \subseteq V$ such that $|S_{\textsc{mini}}| \le k$ and
$|{\cal F}[S_{\textsc{mini}}]| \ge |S_{\textsc{mini}}| + k$.
\end{definition}

\begin{lemma}\label{miniDominatingSet}
A reduced hypergraph $H = (V, {\cal F})$ has a hitting set of size at most $m-k$ if and only if it has a mini-hitting set. Moreover,
\begin{enumerate}
 \item Given a mini-hitting set $S_{\textsc{mini}}$, we can construct a hitting set $S$ with $|S| \le m-k$ such that $S_{\textsc{mini}} \subseteq S$ in polynomial time.
\item  Given a hitting set $S$ with $|S| \le m-k$, we can construct a mini-hitting set $S_{\textsc{mini}}$ such that $S_{\textsc{mini}} \subseteq S$ in polynomial time.
\end{enumerate}
\end{lemma}

\begin{proof}
 \begin{enumerate}
 \item For each edge $e$ not hit by $S_{\textsc{mini}}$, pick one vertex
in $e$ and add it to $S_{\textsc{mini}}$. The resulting set $S$ contains
at most $m-k$ vertices and hits every edge of ${\cal F}$.
\item
If $|S| \le k$ then $S$ itself is a mini-hitting set.

If $|S| > k$, construct $S_{\textsc{mini}}$ as follows. Let $S_0 =
\emptyset$, and for every $0 \le i \le m-k-1$, let $S_{i+1} =
S_{i}\cup\{v\}$, where $v \in S \backslash S_i$ is picked to maximise
$|{\cal F}[v] \backslash {\cal F}[S_i]|$.
Suppose for a contradiction that $|{\cal F}[S_k]| < |S_k| + k$. Then for
some $j<k$, $|{\cal F}[S_{j+1}]|\le |{\cal F}[S_j]|+ 1$. Thus by
construction, $|{\cal F}[S_{i+1}]| \le |{\cal F}[S_i]| + 1$ for all $i\ge
j$. It follows that $|{\cal F}[S]| = |{\cal F}[S_{m-k}]| < |S_{m-k}| + k =
m$, a contradiction.
Therefore $|{\cal F}[S_k]| \ge |S_k| + k$, and thus $S_k$ is the required
$S_{\textsc{mini}}$.
\end{enumerate}
\end{proof}

We now describe a greedy algorithm which constructs a set $S^* \subseteq V$. Either $S^*$ is a mini-hitting set,  or ${\cal F}[S^*]$ has some useful properties which will allow us to bound $|V|$.

Start with $S^* = \emptyset$. While $|{\cal F}[S^*]| < |S^*| + k$ and there exists $v \in V$ with $|{\cal F}[v] \backslash {\cal F}[S^*]| > 1$, do the following: Pick a vertex $v \in V$ such that  $|{\cal F}[v] \backslash {\cal F}[S^*]|$ is as large as possible, and add $v$ to $S^*$.

If $S^*$ is a mini-hitting set, then by Lemma \ref{miniDominatingSet} we are done. We will now assume that $S^*$ is not a mini-hitting set. Let ${\cal C} = {\cal F}[S^*]$, and let ${\cal I} = {\cal F} \backslash {\cal C}$.

\begin{lemma}\label{CProperties}
Suppose $S^*$ is not a mini-hitting set. Then we have the following:
 \begin{enumerate}
\item $|S^*| < k$.
  \item $|{\cal C}| < 2k$.
\item For all $v \in V$, $|{\cal C}[v]| \ge 1$ and $|{\cal I}[v]| \le 1$.
\item For all $v \in V$, $d(v) \le k$.
 \end{enumerate}
\end{lemma}

\begin{proof}
 \begin{enumerate}
 \item Suppose for a contradiction $|S^*| \ge k$. Then at some point in the construction of $S^*$ we have $|S^*| = k$. Observe that at each stage in the construction of $S^*$, $|{\cal F}[S^*]| \ge 2|S^*|$. It follows that when $|S^*| = k, |{\cal F}[S^*]| \ge |S^*| + k$, and the algorithm stops. Note that $S^*$ is a mini-hitting set, a contradiction.
\item Suppose for a contradiction that $|{\cal C}| \ge 2k$. Then since $|S^*| < k$, $|{\cal F}[S^*]| = |{\cal C}| \ge |S^*| + k$, and so $S^*$ is a mini-hitting set, a contradiction.
\item Since $|S^*| < k$ but $S^*$ is not a mini-hitting set, the construction of $S^*$ must stop because $|{\cal F}[v] \backslash {\cal F}[S^*]| \le 1$ for all $v \in V$, i.e. $|{\cal I}[v]| \le 1$. By Lemma \ref{noIslands2}, $d(v)\ge 2$ for all $v \in V$. Since $d(v) = |{\cal C}[v]| + |{\cal I}[v]|$, it follows that $|{\cal C}[v]| \ge 1$.
\item Suppose for a contradiction that there exists $v \in V$ with $d(v) > k$. Then in the construction of $S^*$, we first add a vertex $u$ to $S^*$ with $d(u) > k$. We therefore have a set $S^*$ with $|S^*| = 1, |{\cal F}[S^*]| \ge k+1 = |S^*| + k$, and so the algorithm terminates and $S^*$ is a mini-hitting set, a contradiction.
\end{enumerate}
\end{proof}

We now have that ${\cal F} = {\cal C} \uplus {\cal I}$, with $|{\cal C}|<2k$, and every vertex in $V$ hits at least one edge in ${\cal C}$ and at most one edge in ${\cal I}$. Furthermore
$2 \le d(v) \le k$ for every $v \in V$, and $|e| \ge 2$ for every $e \in {\cal F}$.
We are no longer interested in $S^*$.

Using ${\cal C}$ and ${\cal I}$, we introduce another reduction rule that will bound $|V|$ and consequently $|{\cal F}|$.

\begin{krule}\label{Cneighbourhood} Reduce $(H,k)$ using Rules \ref{subset2}, \ref{subelement2} and
\ref{trivial2}, and let ${\cal C}$ be as defined above.
For any ${\cal C}' \subseteq {\cal C}$, let $V({\cal C}') = \{ v \in V: {\cal C}[v] = {\cal C}' \}$.
If $|V({\cal C}')| > k$, pick a vertex $v \in V({\cal C}')$ and  set $H:=H-v$.
\end{krule}


\begin{lemma}\label{nreduction}
 Let $(H, k)$ and $(H', k)$ be instances of \HSmk{} such that $(H, k)$ is reduced under Rules \ref{subset2}, \ref{subelement2} and \ref{trivial2} and $(H', k)$ is derived from $(H, k)$ by an application of Rule \ref{Cneighbourhood}. Then $(H', k)$ is a {\sc Yes}-instance if and only if $(H, k)$ is a {\sc Yes}-instance.
\end{lemma}

\begin{proof}
 Let $v$ be the vertex removed from $V$ during an application of Rule \ref{Cneighbourhood}. By Lemma \ref{miniDominatingSet}, $(H, k)$ is a {\sc Yes}-instance if and only if $H$ has a mini-hitting set. It is therefore enough to show that $H$ has a mini-hitting set if and only if $H - v $ has a mini-hitting set.

Suppose $S_{\textsc{mini}} \subseteq  V$ is a mini-hitting set for $H$, and assume that $v \in S_{\textsc {mini}}$.
By Rule \ref{subelement2}, each $u \in V({\cal C}')$ is in a different edge $e_u \in {\cal I}$.
 Furthermore, by Lemma \ref{CProperties} part 3, $e_u\cap e_{u'} = \emptyset$ for any $u \neq u' \in V({\cal C}')$.
 As $|V({\cal C}')|>k$, it follows that there exists $u \in V({\cal C}')$ such that $e_u \cap S_{\textsc {mini}} = \emptyset$, i.e. $e_u \notin {\cal F}[S_{\textsc {mini}}]$. Therefore $S_{\textsc {mini}}' = (S_{\textsc {mini}} \backslash \{ v \} ) \cup \{ u \}$ hits $e_u$, which is not hit by $S_{\textsc {mini}}$, and the only edge which is hit by $S_{\textsc {mini}}$ but not $S_{\textsc {mini}}'$ is $e_v$. Therefore $|S_{\textsc {mini}}'| = |S_{\textsc {mini}}|$ and $|{\cal F}[S_{\textsc {mini}}']| \ge |{\cal F}[S_{\textsc {mini}}]| \ge |S_{\textsc {mini}}| + k =
|S_{\textsc {mini}}'| + k$, so $S_{\textsc {mini}}'$ is a mini-hitting set that does not contain $v$. Therefore
 $S_{\textsc {mini}}'$ is a mini-hitting set for $H - v $.

The reverse direction is trivial: If $S_{\textsc {mini}}$ is a mini-hitting set for $H - v$ then it
 is also a mini-hitting set for $H$.
\end{proof}

Note that although the number of subsets of ${\cal C}$ can be exponential in $k$, Rule \ref{CProperties} can be run in polynomial time. This is because we do not need to check every subset ${\cal C}' \subseteq {\cal C}$; it is enough to calculate ${\cal C}[v]$ for each $v \in V$ and only calculate $|V({\cal C}')|$ if there exists $v \in V$ for which ${\cal C}[v] = {\cal C}'$.

\begin{theorem} \label{mKernel}
 \HSmk{} has a kernel with at most $k4^k$ vertices and at most $k4^k$ edges.
\end{theorem}

\begin{proof} Let $(H,k)$ be an instance of \HSmk{} irreducible by the above four reduction rules and let $H=(V,{\cal F}).$
The number of possible subsets ${\cal C}' \subseteq {\cal C}$ is $2^{|{\cal C}|} < 2^{2k}$. Therefore by Rule \ref{Cneighbourhood} $n=|V| < k2^{2k} = k4^k$.

To bound $m=|{\cal F}|$ recall that $d(v) \le k$ for all $v \in V$, and $|e| \ge 2$ for all $e \in {\cal F}$. It follows that $|{\cal F}| \le k|V|/2 < k^22^{2k-1}$. We can improve this bound as follows.

Here we make use of Theorem \ref{thm:n}, which we prove in the next
section. If $m \le n$ then obviously $m \le k4^k$. Suppose that $m >n$. Then finding a hitting set of size $m-k$ is equivalent to finding a hitting set of size $n-k'$, where $k' = (n-m) + k$ is less than $k$. By Theorem \ref{thm:n}, this has a kernel with $ m\le d(d+1)k' < d(d+1)k$, where $d$ is the degeneracy of $H$. By Lemma \ref{Cneighbourhood} part 4, we have $d \le k$ and, thus, $m \le k^2(k+1)$. Therefore we have a kernel with $m \le k2^{2k}$ in either case.
\end{proof}

We now show that our exponential kernel for \HSmk{} cannot be improved to a
polynomial size one, given certain complexity assumptions. We make use of a result of
Dom et al. \cite{DLS09} who proved the following:

\begin{theorem}\label{m+kNoKernel}
 \HSmkk{} does not have a polynomial kernel, unless $coNP \subseteq NP/poly$.
\end{theorem}

We may now prove the following theorem:

\begin{theorem}\label{thm:nopolyk}
 \HSmk{} does not have a polynomial kernel, unless $coNP \subseteq NP/poly$.
\end{theorem}

\begin{proof}

Assume that \HSmk{} has a polynomial kernel. We will show that \HSmkk{}
has a polynomial kernel,
a contradiction unless $coNP \subseteq NP/poly$.

Consider an instance of \HSmkk{}, in which we are given a hypergraph $H =
(V, {\cal F})$ with $|V| = n$,
 $|{\cal F}| = m$, together with an integer $p$, and are asked to find a
hitting set in $H$ of size $p$.
Let $k = m-p$. Observe that $H$ has a hitting set of size $p$ if and only
if $H$ has a hitting set of size $m-k$,
and therefore we can view our instance of \HSmkk{} as an instance of \HSmk.
By our assumption, there is a transformation which produces a hypergraph
$H' = (V', {\cal F}')$ with
$|V'| = n'$, $|{\cal F}'| = m'$ together with an integer $k'$, such that
$H$ has a hitting set of size
$m-k$ if and only if $H'$ has a hitting set of size $m'-k'$. Furthermore,
$m', n' \le P(k)$ for some
polynomial $P$, and the transformation takes time polynomial in $n$ and
$m$. We may assume without
loss of generality that $P$ is an increasing function.

Let $p' =  m' - k'$, and observe that $H$ has a hitting set of size $p$ if
and only if $H'$ has a hitting set of size $m'+p'$.
Therefore $H'$ with parameter $p'$ is an equivalent instance of \HSmkk{}
which can be constructed in time polynomial in $m$ and $n$, and $m', n'
\le P(k) = P(m-p) \le P(m+p)$, i.e. the size of the instance is bounded by
a polynomial in the original parameter. It remains to show that the new
parameter $m'+p'$ is also bounded by a function of the original parameter,
but this follows from the fact that $p'\le m'$.
\end{proof}

\section{Hitting Set parameterized below $n$}\label{sec:n}

 Unlike  \HSmk,  \HSnk{} is not fixed-parameter tractable unless $FPT=W[1].$

\begin{theorem}
 \HSnk{} is $W[1]$-complete.
\end{theorem}

\begin{proof}

To show hardness, we use a well-known reduction from
\textsc{independent set}, in which we are given a graph $G=(V,E)$ and
are asked whether it contains an independent set $V' \subseteq V$ set
of size $k$, where $k$ is the parameter.  In our instance of \HSnk, we
let $H$ be $G$ viewed as a hypergraph.
 Then for any $V' \subseteq V$ with $|V'|
= k$, $V'$ is an independent set in the graph if and only if every
edge contains a member of $V\backslash V'$, i.e. $V\backslash V'$ is a
hitting set.

To show membership in $W[1]$, we reduce\HSnk{} to the problem
$p\textsc{-WSat}(\Gamma_{2,1}^-)$, described in Flum and Grohe
\cite{FlumGrohe06}.
$\Gamma_{2,1}^-$ is the class of CNF formulas which contain only
negative literals.
In the parameterized problem $p\textsc{-WSat}(\Gamma_{2,1}^-)$, we are
given a formula in $\Gamma_{2,1}^-$ and an integer parameter $k$, and
we are asked whether the formula has a satisfying assignment in which
exactly $k$ variables are assigned {\sc True}.
It follows from Theorem 7.29 in Flum and Grohe \cite{FlumGrohe06} that
$p\textsc{-WSat}(\Gamma_{2,1}^-)$ is in $W[1]$ (a more general problem
is in  $W[1]$).

For an instance of \HSnk, let $V = \{ v_1, \dots v_n\}$ be the vertices
 and ${\cal F} = \{ e_1, \dots e_m\}$ the edges in $H$. For each edge $e \in {\cal F}$, we let the clause
$C_e = \bigvee_{v_i \in e}\bar{x_i}$, and let our formula be
$\bigwedge_{j \in [m]} C_{e_j}$.
 Then there is a hitting set of size $(n - k)$ if and only if the
formula has a satisfying assignment in which exactly $k$ variables are
assigned {\sc True}. This is precisely the problem
$p\textsc{-WSat}(\Gamma_{2,1}^-)$, and so we are done.
 \end{proof}

Note that in the hardness proof above, every set in the \HSnk{} instance was of size 2. This means that \HSnk{} is $W[1]$-hard even for the subcase where the edge size is bounded by $r$, for any $r \ge 2$.
Therefore if we let the parameter be $k + \max_{e \in {\cal F}}|e|$, the problem is still $W[1]$-hard.

Another approach would be to consider the degree of the vertices as an additional parameter.
Under this parameterization the problem does turn out to be fixed-parameter tractable;
in fact we prove a stronger result by showing that the problem is fixed-parameter tractable with respect
to $k + d$, where $d$ is the degeneracy of $H$.
This is the problem \HSnkp.

We begin with the following simple result on the chromatic number of a $d$-degenerate hypergraph.
For a hypergraph $H = (V, {\cal F})$, a mapping $c:\ V\rightarrow [t]$ is called a {\em proper $t$-coloring} if
each edge $e$ of $H$ of cardinality at least 2 is not {\em monochromatic}, i.e., $e$ has vertices $u,v$ such that $c(u)\neq c(v)$.
Here $c(u)$ is the {\em color} of $u.$
The {\em chromatic number} $\chi(H)$ of a hypergraph $H$ is the minimum integer $t$ for which $H$ has a proper $t$-coloring.

\begin{lemma}\label{lem:chrom}
The chromatic number of a $d$-degenerate hypergraph is at most $d+1.$
\end{lemma}
\begin{proof}
The proof is by induction on the number $n$ of vertices of $H.$ If $n=1$ then $H$ has no edge of cardinality 1, and
so $\chi(H)=1.$ Now assume that $n\ge 2.$
Let $v$ be a vertex of minimum degree $q$ in $H = (V, {\cal F})$.  By the induction
hypothesis and definition of a $d$-degenerate hypergraph, $\chi(H\circleddash \{v\})\le d+1$.
Consider a $(d+1)$-coloring of $H\circleddash \{v\}$ and
edges $e_1,\ldots ,e_q$ of cardinality at least 2 containing $v$. Note that $q\le d$ and form
a set $C$ of colors by
picking one color used in each $e_i$ (if any vertex in $e_i$ is colored). If
$C$ is empty, add to it color 1. Clearly, $|C|\le d$ and, thus, there is a
color $t$ not in $C$ among colors in $[d+1]$. Assign $v$ using color $t$ and use one
of the colors in $C$ to color all other uncolored vertices. Observe that
none of $e_1,\ldots ,e_q$ is monochromatic.
\end{proof}

To get rid of edges of cardinality 1, we use the following rule whose correctness is easy to see.

\begin{krule}\label{trivial}
 If there exist $v \in V$, $e \in {\cal F}$ such that $e = \{ v \}$, then  replace $H=(V,{\cal F})$ by
$H\circleddash \{v\}$.
Keep $k$ the same.
\end{krule}

For a hypergraph $H$, a set $S$ of vertices is {\em independent} if $S$ does not contain any edge of $H.$

\begin{theorem}\label{thm:n}
The problem \HSnkp{} admits a kernel with less than $(d+1)k$ vertices and $d(d+1)k$ edges.
\end{theorem}
\begin{proof}
Let $H$ be a $d$-degenerate hypergraph. Using Rule \ref{trivial} as long as possible, we reduce $H$ to
a $d$-degenerate hypergraph with no edge of cardinality 1. By Lemma \ref{lem:chrom}, $\chi(H)\le d+1$.
Consider a proper $\chi(H)$-coloring of $H$ and a
largest set $S$ of vertices of $H$ assigned the same color. Clearly, $|S|\ge |V|/(d+1)$.

Now observe that $T$ is a hitting set of $H=(V,{\cal F})$ if and only if $V\setminus T$ is an independent set.
Thus, if $|V|/(d+1)\ge k$, the answer to \HSnkp{} is {\sc Yes}. Otherwise, $|V|<(d+1)k$.

To prove that $|{\cal F}|<d(d+1)k$, choose a vertex $v$ of minimum
degree and observe that $d(v) \le d$. Now delete $v$ from $V$ and
${\cal F}[v]$ from ${\cal F}$, and choose a vertex $v$ of minimum degree
again,
and  observe that $d(v) \le d$. Continuing this
procedure we will delete all edges in ${\cal F}$ and thus $|{\cal F}|\le
d|V|<d(d+1)k$.
\end{proof}

In a directed graph $G=(V, A)$, a \emph{dominating set} is a set
$V' \subseteq V$ such that for every vertex $u \in V \backslash V'$,
there is a vertex $v \in V'$ such that there is an arc from $v$ to $u$.
Recall that in \textsc{Directed Nonblocker}, we are given a directed graph $G$ with $n$
vertices and an integer $k$, and asked
whether $G$ has a dominating set with at most $n-k$ vertices.

\begin{corollary}\label{cor:dn}
\textsc{Directed Nonblocker} has a kernel with at most $k^2+k-1$ vertices.
\end{corollary}

\begin{proof}
 Let $(G=(V, A), k)$ be an instance of \textsc{Directed Nonblocker} with $|V|=n$. If $G$ has a vertex $v$ of
out-degree at least $k$,
then $V\setminus \{w\in V:\ vw\in A\}$ is a dominating set of size at most $n-k.$ Thus, we may assume that
the maximum out-degree of $G$ is at most $k-1$.

We construct an instance of {\HSnk} as follows. Let $H = (V, {\cal F})$, where ${\cal F} = \{ N^-[v] : v \in V \}$,
$N^-[v]=\{v\}\cup \{u\in V:\ uv\in A\}.$
Observe that $N^-[v]$ is hit by a set $S \subseteq V$ if and only if $v \in S$ or $v$ is
dominated by a vertex in $S$. Therefore, $H$ has a hitting set of size $|{\cal F}| - k = |V|-k$ if and
only if $G$ has a dominating set of size $|V|-k$.

Since the maximum out-degree of $G$ is at most $k-1$, the maximum degree of a vertex
in $H$ is at most $k$. Thus, the degeneracy $d$ of $H$ is at most $k$
and the result follows from Theorem \ref{thm:n}.
\end{proof}

\section{Directed Nonblocker}\label{sec:nb}

In this section, we improve the bound of Corollary \ref{cor:dn} for $k>2.$

It is well-known that every hypergraph $H=(V,{\cal F})$ in which each edge has at least two vertices,
has a hitting set of cardinality at most $(|V|+|{\cal F}|)/3$, cf. \cite{TY2007}.
We start from a minor extension of this result.

\begin{lemma}\label{lem:hh}
Let $H=(V,{\cal F})$ be a hypergraph such that every edge has at least two vertices apart from, possibly, one edge that has just one vertex.
If $H$ has a one-vertex edge $e=\{v\}$, let there be another edge $f$ of $H$ containing $v.$ Then $H$ has a hitting set of cardinality at most $(|V|+|{\cal F}|)/3.$
\end{lemma}
\begin{proof}
Let $n=|V|$ and $m=|{\cal F}|.$ The proof is by induction on $n\ge 2.$ If $n=2$, then $H$ has a hitting set of cardinality 1 and $n+m\ge 3.$
Now assume that $n\ge 3$ and let $t(H)$ be the minimum cardinality of a hitting set in $H.$ If $H$ has a one-vertex edge $e=\{v\}$, then set $u=v$.
Otherwise, let $u$ be a vertex of $H$ of maximum degree. Remove $u$ from $H$ together with all edges containing $u$ and
all vertices contained only in the removed edges.
Denote the resulting hypergraph by $H'$ and let $n'$ and $m'$ be the number of vertices and edges, respectively, in $H'.$ Then $3t(H)\le 3+3t(H')\le 3 + n'+m'\le n+m.$ The second inequality in this chain of inequalities is by the induction hypothesis and the third inequality is due to the fact that either we remove at least two edges and one vertex or at least two vertices and one edge.
\end{proof}

For a digraph $D$, let $\gamma(D)$ denote the minimum size of a dominating set in $D.$
Using Lemma \ref{lem:hh}, it is easy to prove the following key lemma of this section.

\begin{lemma}\label{lem:2/3}
Let $D$ a digraph on $n$ vertices, none of which are isolated, and let $D$ have
at most one vertex of in-degree zero. Then $\gamma(D)\le 2n/3.$
\end{lemma}
\begin{proof}
We construct an instance $H=(V,{\cal F})$ of {\HSnk} as in the proof of Corollary \ref{cor:dn}. The lemma follows from that facts that $H$ satisfies the conditions of Lemma \ref{lem:hh}, $|V|=|{\cal F}|$, and the minimum cardinalities of a hitting set in $H$ and a dominating set in $D$ coincide.
\end{proof}

\begin{theorem}
\textsc{Directed Nonblocker} has a kernel with at most $3k-1$ vertices.
\end{theorem}
\begin{proof}
Let $D$ be a digraph with $n$ vertices. If $D$ has isolated vertices, then delete them without changing the answer to \textsc{Directed Nonblocker}
as all of them must be in any dominating set of $D$. Thus, we may assume that $D$ has no isolated vertices.
Let $S$ be the set of all vertices of $D$ of in-degree zero. Assume that $|S|>1$. Then contract all vertices of $S$ into one vertex $s$ which dominates all vertices dominated by $S$. Let $D'$ be the resulting digraph. Since all vertices
of $S$ must be in any dominating set of $D$, the answers to \textsc{Directed Nonblocker} on $D$ and on $D'$ are the same. Thus, we may assume
that $|S|\le 1$. Then, by Lemma \ref{lem:2/3}, $\gamma(D)\le 2n/3$ and, thus, if $n-k\ge 2n/3$, the answer to \textsc{Directed Nonblocker} is {\sc Yes}.
Otherwise, $n-k< 2n/3$ and $n\le 3k-1.$
\end{proof}

To obtain a smaller kernel for \textsc{Directed Nonblocker}, it might be helpful to use further results on
hitting sets of hypergraphs with a lower bound on the minimum size of an edge. Chv\'{a}tal and McDiarmid
\cite{CM1992}  and Tuza \cite{Tuza1990} proved independently that a hypergraph $H=(V,{\cal F})$
with minimum edge size equal three, has a
hitting set of size at most $(|V|+|{\cal F}|)/4.$ Thomass\'{e} and Yeo \cite{TY2007} showed that if the minimum
edge in a hypergraph $H=(V,{\cal F})$ is four and the minimum size of a hitting set of $H$ is $t$,
then $21t\le 5|V|+4|{\cal F}|.$

\section{Further Research}\label{sec:fr}

For a hypergraph $H$, let $\alpha(H)$ be the maximum size of an independent set of $H$. In the proof of Theorem \ref{thm:n}, we observed that \HSnkp{} is equivalent to problem of deciding whether $\alpha(H)\ge k$ for a $d$-degenerate hypergraph $H$. This observation and the inequality $\alpha(H)\ge n/(d+1)$, where $n$ is the number of vertices in $H$, have allowed us to obtain a linear kernel for \HSnkp. However, the inequality $\alpha(H)\ge n/(d+1)$ suggests that, in fact, to have the parameter small in relevant cases (as it should be in the spirit of parameterized algorithmics) it makes more sense to consider the following parameterization above tight lower bound: decide whether for a $d$-degenerate hypergraph $H$ we have $\alpha(H)\ge n/(d+1)+\kappa$, where $\kappa$ is the new parameter. (Problems parameterized above tight lower bounds were studied in several papers including \cite{GKLM2010,GKMY2010,GKSY2010,Mahajan99,MahajanRamanSikdar09}.)

It would be interesting to determine the parameterized complexity of the last problem even in the case of graphs. To the best of our knowledge, the only related
result was obtained by Gutin et al. \cite{GKLM2010} who observed that the last problem has a linear kernel for graphs of maximum degree at most $d.$
Indeed, in the case of graphs with maximum degree at most $d$, the existence of a linear kernel is an easy consequence of the following Brooks' Theorem \cite{West01}: for a graph $G$ with maximum degree at most $d$ we have $\chi(G)\le d$ unless one of the connectivity components of $G$
is $K_{d+1}$ or, if $d=2$ and one of the connectivity components of $G$ is an odd cycle. Brook's Theorem was extended to hypergraphs by Kostochka et al. \cite{kost} who proved that for a connected hypergraph $H$ with all edges of cardinality at least 2 and of maximum degree at most $d$ we have $\chi(H)\le d$ unless $H$ has only one edge of cardinality at least 3 or $H$ is a graph (in which case the odd cycle and complete cases apply). Thus, the observation of \cite{GKLM2010} can be extended to hypergraphs.

\end{document}